\documentclass[twocolumn,reprint,aps,prl]{revtex4-1}
\usepackage[latin9]{inputenc}
\usepackage{xcolor}
\usepackage{amsmath}
\usepackage{amssymb}
\usepackage{graphicx}
\PassOptionsToPackage{normalem}{ulem}
\usepackage{ulem}
\usepackage[unicode=true,
 bookmarks=false,
 breaklinks=false,pdfborder={0 0 1},backref=section,colorlinks=true]
 {hyperref}
\hypersetup{
 linkcolor=blue,anchorcolor=blue,citecolor=blue,urlcolor=blue}
\usepackage{breakurl}

\makeatletter

\providecolor{lyxadded}{rgb}{0,0,1}
\providecolor{lyxdeleted}{rgb}{1,0,0}

\usepackage{dcolumn}
\usepackage{bm}
\usepackage{times}
\usepackage{multirow}
\usepackage{amsfonts}

\makeatother

\begin{document}

\title{Intrinsic Magnetoresistance in Three-Dimensional Dirac Materials}

\author{Huan-Wen Wang}

\affiliation{Department of Physics, The University of Hong Kong, Pokfulam Road,
Hong Kong, China}

\author{Bo Fu}

\affiliation{Department of Physics, The University of Hong Kong, Pokfulam Road,
Hong Kong, China}

\author{Shun-Qing Shen}
\email{sshen@hku.hk}

\affiliation{Department of Physics, The University of Hong Kong, Pokfulam Road,
Hong Kong, China}

\date{24 March, 2018}
\begin{abstract}
Recently, negative longitudinal and positive in-plane transverse magnetoresistance
have been observed in most topological Dirac/Weyl semimetals, and
some other topological materials. Here we present a quantum theory
of intrinsic magnetoresistance for three-dimensional Dirac fermions
at a finite and uniform magnetic field $B$. In a semiclassical regime,
it is shown that the longitudinal magnetoresistance is negative and
quadratic of a weak field $B$ while the in-plane transverse magnetoresistance
is positive and quadratic of $B$. The relative magnetoresistance
is inversely quartic of the Fermi wave vector and only determined
by the density of charge carriers, irrelevant to the external scatterings
in the weak scattering limit. This intrinsic anisotropic magnetoresistance
is measurable in systems with lower carrier density and high mobility.
In the quantum oscillation regime a formula for the phase shift in
Shubnikov-de Hass oscillation is present as a function of the mobility
and the magnetic field, which is useful for experimental data analysis.
\end{abstract}
\maketitle
\emph{Introduction}-Magnetoresistance is the value change of electric
resistance of a material in an applied magnetic field, and depends
on the mutual orientation of the electric current and the magnetic
field. In a sufficient weak field, the origin of the magnetoresistance
is highly related to the Lorentz force experienced by charge carries
in the magnetic field and the spin-dependent scattering of electrons
\cite{Abrikosov-book,Pippard1989a}. Recently a positive in-plane
transverse and negative longitudinal magnetoresistance have been observed
in topological Dirac and Weyl semimetals\cite{Armitage2018,Lu2017,Kim2013,Xiong2015,Li2016,Li2015a,Li2015,Huang2015,Zhang2016a,Wang2016a,Zhang2017},
and some other metallic materials \cite{Wang2012,He2013,Wiedmann2016,Assaf2017}.
Especially the negative longitudinal magnetoresistance in Dirac and
Weyl semimetals attracts great interests as its physical origin is
possibly related to the chiral anomaly\cite{Adler1969,Bell1969,Nielsen1983},
a purely quantum mechanical effect, of the three-dimensional Weyl
fermions in the electric and magnetic fields \cite{Son2013,Burkov2014a,Burkov2015,Goswami2015,Gorbar2014}.
Several mechanisms without chiral anomaly are also proposed for conventional
and topological metals \cite{Gao2017,Andreev2018}. 

On the other hand, while the touching points of conduction and valence
bands in the Weyl semimetals are protected topologically, the Dirac
semimetals are located between conventional and topological insulators
\cite{Murakami2007,Burkov2016,Yan2017}. A small lattice distortion
or external field can open a small energy gap in the band structure.
Furthermore narrow gap semiconductors are also well described by the
Kane model \cite{Kane1957} in which the conduction and valence bands
are strongly coupled together. A class of the gapless topological
semimetals, and narrow-gap semiconductors or topological materials
can be well described by an effective multi-band Dirac model \cite{Shen2012,Shen-11spin,Zawadzki-17jpcm}.
In this class of materials, when the Fermi energy is located above
the bottom of the conduction band, the transport properties are also
affected by the existence of the valence bands as well as the conduction
band. The strong band coupling in these materials produce prosperous
physics of Berry phase in electron dynamics \cite{Chang2008,Xiao2010}. 

In this Letter we propose an intrinsic origin of magnetoresistance
of three-dimensional Dirac fermions in a finite magnetic field in
the framework of the Kubo formula with the help of Landau levels.
In the semiclassical regime, the quadratic corrections of a magnetic
field are found to both longitudinal and in-plane transverse resistivity
and the electrical mobility. As a consequence the relative magnetoresistivity
is quartic of the ratio of the Fermi wave length (the reciprocal of
the Fermi wave vector) to the magnetic length. In the weak scattering
limit the magnetoresistivity is only determined by the carrier density,
and irrelevant to the external scatterings. Thus we dub it the intrinsic
magnetoresistivity. The effect becomes measurable when the Fermi wave
length is comparable with the magnetic length, i.e., the carrier density
is low such that the Fermi level crosses near the Weyl nodes for the
Dirac semimetals and is close to the bottom of the conduction bands
for the narrow-gap semiconductors or topological insulators. In the
quantum oscillatory regime, a formula for the phase shift is presented
as a function of the mobility and the magnetic field, which will be
useful for data analysis

\begin{figure}
\raggedright{}\includegraphics[scale=0.5]{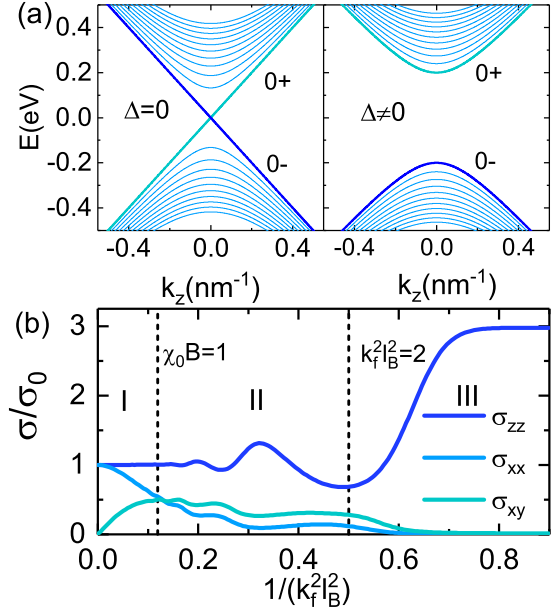}\caption{(a) The band structure of Dirac fermions for the gapless case $\Delta=0$
(the left panel) and the massive case $\Delta\protect\ne0$ (the right
panel). (b) The conductivity as a function of a magnetic field, or
magnetoconductivity of massless Dirac fermions. $\sigma_{zz}$ is
the longitudinal magnetoconductivity for $\mathbf{B}\parallel\mathbf{E}$
configuration, $\sigma_{xx}$ is the in-plane transverse magnetoconductivity
for $\mathbf{B}\perp\mathbf{E}$ configuration, and $\sigma_{xy}$
is the Hall conductivity. The shown dimensionless magnetic field scales
(vertical dashed lines) indicate the borders of the different regimes:
(I) the semiclassical regime ($\chi_{0}B<1$) where Landau levels
are smeared out due to the disorder broadening and the background
dominates the magnetoconductivity; (II) quantum oscillation regime
($\chi_{0}B>1$) where Shubnikov-de Haas oscillation occurs; (III)
the quantum limit regime ($k_{f}l_{B}<\sqrt{2}$), where only zero
Landau level contributes.\label{fig:magnetoconductivity-in-different}}
\end{figure}

\textit{Model and the Kubo-Streda formula for conductivity-}To illustrate
the effect of the intrinsic magnetoresistivity, we start with the
Dirac Hamiltonian in a finite magnetic field, which describes either
the Dirac semimetals or the narrow-gap semiconductors and topological
materials, 
\begin{equation}
\mathcal{H}=\begin{bmatrix}\Delta & v\hbar\mathbf{\sigma}\cdot(\mathbf{k}-e\mathbf{A})\\
v\hbar\mathbf{\sigma}\cdot(\mathbf{k}-e\mathbf{A}) & -\Delta
\end{bmatrix}.\label{Hamiltonian}
\end{equation}
Here $v$ is the effective velocity and $2\Delta$ is the energy gap
between the conduction band and valence band. $\sigma_{\alpha}$ ($\alpha=x,y,z$)
are the Pauli matrices. Without loss of generality, we assume the
magnetic field is applied along the $z-$direction. The vector potential
is then chosen as $\mathbf{A}=(-By,0,0)$. We focus on the situation
in which the Fermi level $\mu$ is above the energy gap $2\Delta$.
In the absence of a magnetic field, the Fermi level $\mu$ is related
to the Fermi wave vector $k_{f}$, $\mu^{2}=\Delta^{2}+(\hbar vk_{f})^{2}$
, or the Fermi wave length $1/k_{f}$. In a finite field, the energy
spectrum has the form, $\varepsilon_{n}^{\zeta}=\zeta\sqrt{v^{2}\hbar^{2}k_{z}^{2}+2n\left(\hbar v/l_{B}\right)^{2}+\Delta^{2}}$
where $l_{B}=\sqrt{\hbar/eB}$ is the magnetic length and $\zeta=\pm1$
is the band index and each band is doubly degenerate in energy for
$n=1,2,\cdots$ , and non-degenerate for $n=0$, as shown in Fig.\ref{fig:magnetoconductivity-in-different}(a).

We consider the short-range point-like impurities $U=u_{0}\sum_{l}\delta(\mathbf{r}-\mathbf{R}_{l})$
with the impurity concentration $n_{i}$. In this work, we utilize
the Kubo-Streda formula\cite{Streda1982} to calculate the matrix
element of conductivity tensor
\begin{align}
\sigma_{\alpha\beta}= & \frac{\hbar e^{2}}{2\pi V}\sum_{k}\int_{-\infty}^{+\infty}d\xi n_{F}(\xi)\text{Tr}[\hat{v}^{\alpha}\frac{dG^{R}}{d\xi}\hat{v}^{\beta}(G^{A}-G^{R})\nonumber \\
 & -\hat{v}^{\alpha}(G^{A}-G^{R})\hat{v}^{\beta}\frac{dG^{A}}{d\xi}]\label{eq:Kubo-streda formula}
\end{align}
here $V$ is the volume of the system, $\hat{v}^{\alpha}\equiv\frac{1}{\hbar}\frac{\partial\mathcal{H}}{\partial k_{\alpha}}$
is the velocity operator along $\alpha-$direction with $\alpha=x,y,z$,
$n_{F}(\xi)=[1+\exp(\frac{\xi-\mu}{k_{B}T})]^{-1}$ is the Fermi-Dirac
distribution with $k_{B}$ being the Boltzmann constant and T being
the absolute temperature, $G^{R/A}(\xi)=\frac{1}{\xi-\mathcal{H}\pm i\gamma}$
are the retarded and advanced Green's functions. In the Born approximation,
the scattering time $\tau=\frac{\hbar}{2\gamma}=\hbar/(2\pi N_{f}n_{i}u_{0}^{2})$
with the density states $N_{f}=\mu k_{f}/(\pi\hbar^{3}v^{3})$ at
the Fermi level. With the help of the eigen functions of the Landau
levels, all the elements of the conductivity tensor can be expressed
as the series summation over the Landau index $n$ at the zero temperature
(see Eq. {[}13,14,19{]} in the Supplementary Material \cite{Note-on-SM}). 

The calculated longitudinal (the electric field is parallel to the
magnetic field) conductivity $\sigma_{zz}$, in-plane transverse (the
electric field is perpendicular to the field) conductivity $\sigma_{xx}=\sigma_{yy}$
and the Hall conductivity $\sigma_{xy}$ are plotted in Fig.\ref{fig:magnetoconductivity-in-different}b,
which can be divided into three different regimes: (I) the semiclassical
regime, (II) the quantum oscillation regime, and (III) the quantum
limit regime. In the semiclassical regime, the energy band broadening
width $\gamma$ is larger than the energy spacing of two adjacent
Landau levels near the Fermi level $\mu$, $\gamma>(\varepsilon_{m+1}^{+}-\varepsilon_{m}^{+})/2\approx\hbar^{2}v^{2}eB/(2\hbar\mu)$
or $\chi_{0}B<1$ with the mobility $\chi_{0}=e\hbar v^{2}/(2\gamma\mu)$.
Thus the Shubnikov-de Haas oscillations will be smeared out by disorder
effect in this regime. In the quantum oscillation regime $\chi_{0}B>1$
, the Landau levels near the Fermi level $\mu$ will be well separated
from each other and the quantum oscillations become distinct. Further
increasing the magnetic field $k_{f}l_{B}<\sqrt{2}$, all the charge
carriers will be confined into the lowest Landau level, which is also
named as the quantum limit.

\textit{Intrinsic magnetoresistivity-}In the semiclassical regime,
the longitudinal magnetoconductivity $\sigma_{zz}$ is usually thought
to be absent in the approximation of a spherical Fermi surface. In
the weak field limit we find that $\sigma_{zz}=\sigma_{0}=\frac{e^{2}v^{2}k_{f}^{3}}{3\pi^{2}\mu}\tau$
and the electric mobility $\chi=\chi_{0}=\frac{ev^{2}}{\mu}\tau$,
which are identical to the results of the free Dirac fermions in the
absence of the magnetic field. The in-plane transverse conductivity
decays with increasing magnetic field due to the Lorentz force, $\sigma_{xx}=\frac{\sigma_{0}}{1+\left(\chi_{0}B\right)^{2}}$.
In this case, although the transverse conductivity decays with the
magnetic field, both the longitudinal and transverse magnetoresistivity
are absent, $\rho_{xx}=\rho_{zz}=1/\sigma_{0}$ \cite{Pippard1989a}.
However, a detailed calculation of the series summation of the conductivity
tensor at a finite field shows a quantum correction to either the
conductivity or the mobility. We perform the summation over the Landau
levels with the help of the Hurwitz zeta function $\zeta(s,z)=\sum_{n}\left(n+z\right)^{-s}$
and the digamma function $\psi(z)$, and then utilize the asymptotic
expansion of the digamma function and Hurwitz zeta function for a
large z, $\psi(z)=\log z-\frac{1}{2z}-\frac{1}{12z^{2}}+\cdots$ and
$\zeta(2,z)=\frac{1}{z}+\frac{1}{2z^{2}}+\frac{1}{6z^{3}}+\cdots$,
keeping up to the $\left(k_{f}l_{B}\right)^{-4}$ terms, to evaluate
the conductivity {[}see Sec.S6 in Ref.\cite{Note-on-SM} for the calculation{]}. 

After some cumbersome but straightforward calculation, we find that
the longitudinal conductivity is expressed as $\sigma_{zz}=\sigma_{0}\left[1-\frac{c_{z}}{(k_{f}l_{B})^{4}}\right]$
and the transverse conductivity as $\sigma_{xx}=\frac{\sigma_{0}}{1+\chi^{2}B^{2}}\left[1-\frac{c_{x}}{(k_{f}l_{B})^{4}}\right]$
with the mobility $\chi=\chi_{0}\left[1+\frac{c_{\chi}}{(k_{f}l_{B})^{4}}\right]$.
The mobility is derived from the ratio of the Hall conductivity to
the transverse conductivity, $\chi=\sigma_{xy}/\sigma_{xx}B$. The
quadratic correction is consistent with the Casimir-Onsager reciprocity
relation $\sigma_{\alpha\alpha}(B)=\sigma_{\alpha\alpha}(-B)$ as
a consequence of the time-reversal symmetry \cite{Onsager1931}. The
dimensionless parameter $1/(k_{f}l_{B})$ can be understood as the
ratio of the Fermi wave length $\lambda_{f}=1/k_{f}$ to the magnetic
length $l_{B}$. The Fermi wave vector $k_{f}$ is determined by the
carrier density $\varrho$, i.e., $k_{f}=\left(3\pi^{2}\varrho\right)^{1/3}$.
Alternatively, $\left(k_{f}l_{B}\right)^{-2}=\frac{B}{2B_{F}}$ with
$B_{F}=\frac{\hbar}{2e}k_{f}^{2}$. Comparison of these semiclassical
formulae and the numerical results are shown in Fig. 2a and 2c for
massless and massive Dirac fermions, respectively. We find that the
semiclassical formulae for conductivity are in a good agreement with
the numerical results in whole semiclassical regime. 

The magnetoresistivity $\rho_{\alpha\alpha}(B)$ is derived from the
inverse of the conductivity tensor. Here we stress the importance
of the complete set of the conductivity tensor to produce the accurate
and correct behaviors of the magnetoresistivity. Denote the relative
magnetoresistivity by $\delta\rho_{\alpha\alpha}=\rho_{\alpha\alpha}(B)/\rho(0)-1$.
In a weak field, the relative magnetoresistivity can be expressed
as 

\begin{equation}
\delta\rho_{\alpha\alpha}(B)=\frac{c_{\alpha}}{\left(l_{B}k_{f}\right)^{4}}=c_{\alpha}\left(\frac{B}{2B_{F}}\right)^{2}.
\end{equation}
The formula is also in a good agreement with the numerical results
as shown in Fig. 2b and Fig. 2d. The relative magnetoresistivity is
the main result in this work.

\begin{figure}
\includegraphics[scale=0.35]{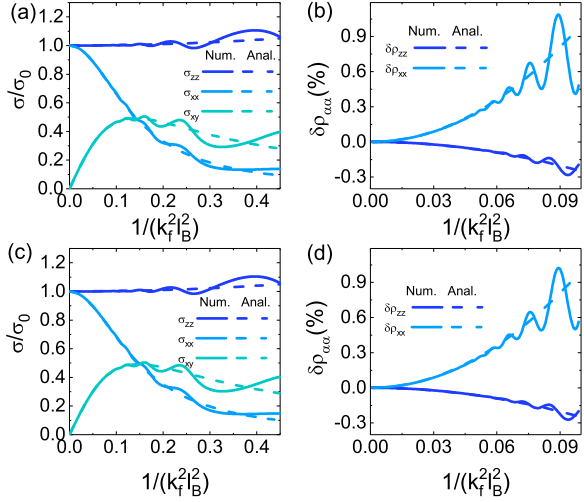}\caption{(a) The magnetoconductivity and (b) magnetoresistivity for massless
Dirac fermions ($\Delta=0$). The dashed lines are the explicit numerical
results and the solid lines are the corresponding analytic results
in the semiclassical regime. (c) The magnetoconductivity and (d) magnetoresistivity
for massless Dirac fermions ($\Delta/\hbar vk_{f}=0.3$ ). The broadening
width is $\frac{\gamma}{\hbar vk_{f}}=0.07$. $k_{f}=0.13\text{nm}^{-1}$
throughout the work. The calculated coefficients are $c_{x}=1$ for
both the massless and massive case. \label{fig:magnetoconductivity-and-magnetor}}
\end{figure}

The dimensionless coefficients $c_{\alpha}$ ($\alpha=x,y,z,\chi$)
are functions of the broadening width $\gamma$ and the energy gap
$\Delta$. It is noted that $c_{z}$ for the longitudinal magnetoresistivity
is always negative and $c_{x}=c_{y}$ for the transverse magnetoresistivity
are always positive for either the massless or massive Dirac fermions.
In the weak scattering limit the band broadening width $\gamma\rightarrow0$,
it is found that $c_{x}=1$, $c_{z}=-1/4$ and $c_{\chi}=-3/4$, irrelevant
to the external scattering {[}see Sec.S6 in Ref.\cite{Note-on-SM}
for the calculation{]}. In this case, the magnetoresistivity is determined
by the Fermi wave vector $k_{f}$. For a specific Fermi level $\mu$
the band gap $\Delta$ can tune the Fermi wave vector via $k_{f}=\sqrt{\mu^{2}-\Delta^{2}}/\hbar v$,
but for a specific carrier density $\varrho$, the Fermi wave vector
is given by $k_{f}=\left(3\pi^{2}\varrho\right)^{1/3}$, irrelevant
to the band gap. Thus the magnetoresistivity is determined by the
electronic band structure and is intrinsic. A similar intrinsic magnetoconductivity
was produced by the Berry curvature of the band structure in the semiclassical
theory \cite{Gao2017}. The intrinsic effect can be suppressed by
the strong impurity scattering. The calculated coefficients as functions
of the broadening width $\gamma$ and the gap $\Delta$ as shown in
Fig. 3. For a weak scattering $\gamma<0.1\hbar vk_{f}$, the coefficients
are rather robust against $\gamma$, but decays quickly to zero for
a large $\gamma$. However for a strong disorder scattering the validity
of the Born approximation is a question. For a large gap, the coefficients
also decay very quickly.

\begin{figure}
\includegraphics[scale=0.35]{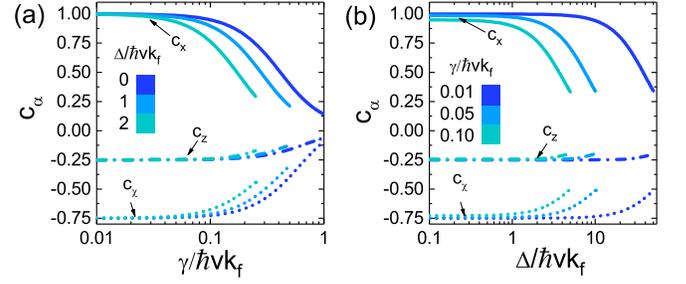}\caption{The dimensionless coefficients ($c_{\alpha}$) of the magnetoresistivities
and electric mobility as a function of (a) the broadening width with
different energy gap and (b) energy gap with different broadening
width. All of the lines are plotted with the constraint of $\hbar^{2}v^{2}k_{f}^{2}>2\Delta\gamma$,
i.e., in the semiclassical regime. \label{fig:The-dimensionless-prefactor}}
\end{figure}

\emph{Phase shift in the quantum oscillation} \emph{regime}-The quantum
oscillation in the regime II is known as the Shubnikov-de Haas oscillation,
which is described by the Lifshitz-Kosevich formula \cite{Shoenberg1984}.
By introducing the Dingle factor $\lambda_{D}=\pi/(\chi_{0}B)$ for
the Lorentz distribution function in the series summation in the conductivity,
which is a function of the mobility $\chi_{0}$ and the magnetic field
$B$, the relative oscillatory part of conductivity is approximately
described by
\begin{equation}
\delta\rho_{\alpha\alpha}^{os}=\frac{d_{\alpha}}{k_{f}l_{B}\cos2\pi\phi(B)}\mathrm{Li}_{\frac{1}{2}}\left(e^{-\frac{\pi}{\chi_{0}B}}\right)\cos[2\pi(\frac{B_{F}}{B}+\phi(B))]
\end{equation}
with the pre-factors $d_{x}=7\sqrt{2}/4$ and $d_{z}=\sqrt{2}$. $\mathrm{Li}_{s}(z)$
is the polylogarithm function of order $s$ and argument $z$. $\phi_{B}$
is not a constant, but a slow-varying phase shift as a function of
the Dingle factor,
\begin{equation}
2\pi\phi(B)=\arctan\left\{ \frac{Re\left[\sqrt{2}\exp(i\frac{3\pi}{4})\mathrm{Li}{}_{\frac{1}{2}}\left(ie^{-\frac{\pi}{\chi_{0}B}}\right)\right]}{\mathrm{Li}{}_{\frac{1}{2}}\left(e^{-\frac{\pi}{\chi_{0}B}}\right)}\right\} .
\end{equation}
In the quantum oscillatory regime, the field $B$ is confined by $\chi_{0}B_{F}>\frac{B_{F}}{B}>1$,
and the value of the Dingle factor is between $\pi/(\chi_{0}B_{F})$
and $\pi$. As a consequence, the phase shift continuously varies
from almost 0 to $-0.238\pi$ as shown in the Fig. 4. For a specific
range of a measurable magnetic field $B$, the value of $\phi(B)$
is mainly determined by the mobility. In the massless case, $\Delta=0$,
usually the mobility can be very large and the factor $\pi/\chi_{0}B_{F}=4\pi\gamma/(v\hbar k_{f})$
is quite small, and the phase shift almost equal to zero for $B$
has the same or less order of $B_{F}$. However for a large gap $\Delta/v\hbar k_{f}\gg1$,
$\pi/\chi_{0}B_{F}=4\pi\gamma\Delta/(v\hbar k_{f})^{2}$ and the phase
shift is close to $-\pi/4$. In practice, $\phi(B)$ and $B_{F}$
can be obtained from the Landau level fan diagram. The phase shift
$\phi(B)$ can be determined by the interpolation line of $n$ versus
$1/B$ (see Fig.1 in Ref. \cite{Note-on-SM}). It is noted that the
phase shift is only a function of the Dingle factor no matter whether
the Dirac fermions are gapless or not. 

\begin{figure}
\includegraphics[width=8cm]{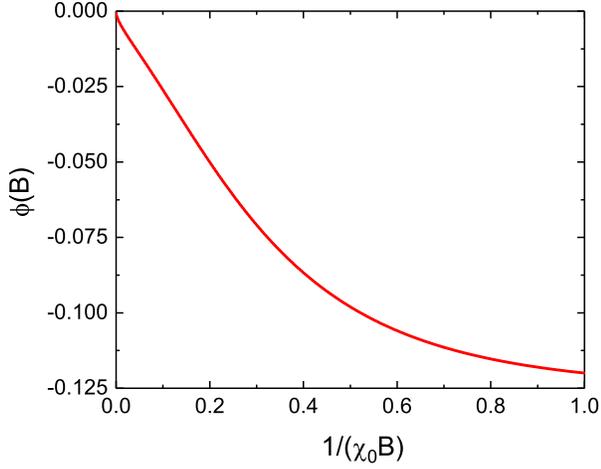}\caption{The phase shift $\phi_{B}$ as a function of $1/(\chi_{0}B)$. \label{fig:Phase shift}}
\end{figure}

\emph{Magnetoconductivity in the quantum limit}-When the magnetic
field grows sufficient large $(k_{f}\ell_{B}\ll1)$, only the Landau
level of $n=0$ is partially filled, i.e., the system is in the quantum
limit regime \cite{Abrikosov1998}. So we only need to consider the
$n=0$ term in Eq.(S13) and Eq.(S14) in Ref. \cite{Note-on-SM}. In
the case the chemical potential varies with the magnetic field as
$\mu=\sqrt{(2\pi^{2}l_{B}^{2}\hbar v\varrho)^{2}+\Delta^{2}}$ and
the scattering time is evaluated as $\tau=\frac{2\pi^{3}\ell_{B}^{4}\hbar^{3}v^{2}\varrho}{n_{i}u_{0}^{2}\sqrt{(2\pi^{2}\ell_{B}^{2}\hbar v\varrho)^{2}+\Delta^{2}}}$
in the Born approximation. The longitudinal and transverse conductivity
satisfy a relation approximately,
\begin{equation}
\sigma_{xx}\sigma_{zz}\simeq\frac{1}{2\pi^{2}l_{B}^{2}}\left(\frac{e^{2}}{h}\right)^{2}.\label{eq:relation}
\end{equation}
The longitudinal conductivity is $\sigma_{zz}=\frac{e^{2}v^{2}\varrho\tau}{\mu}$.
For the massless case of $\Delta=0$, $\tau=\frac{\pi\ell_{B}^{2}\hbar^{2}v}{n_{i}u_{0}^{2}}$
and $\mu=2\pi^{2}\ell_{B}^{2}\hbar v\varrho$. The resulting conductivity
$\sigma_{zz}=\frac{e^{2}\hbar}{2\pi}\frac{v^{2}}{n_{i}u_{0}^{2}}=\text{constant}$
and $\sigma_{xx}\propto B$ which is consistent with the results for
massless Dirac fermions in Ref.\cite{Lu2015,Zhang2016,Klier2017}.
For the large massive case of $\Delta\gg2\pi^{2}\ell_{B}^{2}\hbar v\varrho$,
$\tau=\frac{\pi\ell_{B}^{2}\hbar^{2}v}{n_{i}u_{0}^{2}\Delta}$ and
$\mu\simeq\Delta$. The conductivity$\sigma_{zz}\approx\frac{2\pi^{3}e^{2}\hbar^{3}v^{4}l_{B}^{4}\varrho^{2}}{n_{i}u_{0}^{2}\Delta^{2}}\propto\frac{1}{B^{2}}$,
which indicates a negative magnetoconductivity or positive magnetoresistivity
$\rho_{zz}\propto B^{2}$ in the longitudinal configuration. This
result is consistent with the results for electron gas in semiconductor
with low carrier density\cite{Goswami2015}. Following from Eq. (\ref{eq:relation}),
the corresponding transverse magnetoconductivity is found to be $\sigma_{xx}\propto B^{3}.$

\textit{Discussions}-The negative longitudinal and positive in-plane
transverse magnetoresistivity reflect the anisotropic magneto-transport
in the Dirac materials. The difference of the two resistivities $\rho_{zz}-\rho_{xx}=\frac{c_{z}-c_{x}}{\sigma_{0}}\frac{B^{2}}{\left(2B_{F}\right)^{2}}$
leads to a general relation between the electric field $\mathbf{E}$
and charge current density $\mathbf{j}$ ,

\begin{equation}
\mathbf{E}=\rho_{\perp}\mathbf{j}+\frac{c_{z}-c_{x}}{\sigma_{0}}\frac{\left(\mathbf{j}\cdot\mathbf{B}\right)\mathbf{B}}{\left(2B_{F}\right)^{2}}+\rho_{\perp}\chi\mathbf{B}\times\mathbf{j}.\label{eq:V-I-1}
\end{equation}
with $\rho_{\perp}=\frac{1}{\sigma_{0}}\left(1+c_{x}\frac{B^{2}}{\left(2B_{F}\right)^{2}}\right)$.
In the x-z plane constructed by $\mathbf{B}$ and $\mathbf{j}$ ,
it follows that the resistivity $\rho_{ij}=\rho_{\perp}\delta_{ij}+\frac{c_{z}-c_{x}}{\sigma_{0}}\frac{B_{i}B_{j}}{\left(2B_{F}\right)^{2}}$.
The diagonal resistivity is anisotropic as a function of the angle
$\varphi$ between the magnetic field and electric current density,
\textit{i.e.}, anisotropic magnetoresistivity (AMR), $\rho_{zz}=\frac{1}{\sigma_{0}}\left(1+\frac{c_{z}+c_{x}}{2}\frac{B^{2}}{\left(2B_{F}\right)^{2}}+\frac{c_{z}-c_{x}}{2}\frac{B^{2}}{\left(2B_{F}\right)^{2}}\cos2\varphi\right)$,
and the off-diagonal AMR or planar Hall resistivity is $\rho_{xz}=\frac{c_{z}-c_{x}}{2\sigma_{0}}\frac{B^{2}}{\left(2B_{F}\right)^{2}}\sin2\varphi$.
This planar Hall resistivity satisfies the symmetric relation, $\rho_{xz}=\rho_{zx}$,
unlike the ordinary Hall resistivity which is perpendicular to the
magnetic field, and has an anti-symmetric relation. The oscillatory
amplitude is quadratic in the field $B$. This effect was recently
discussed and explored in the Dirac semimetals \cite{Burkov2017,Nandy2017,Li2017,Wu2017,Kumar2017}. 

The effect becomes strong when the carrier density is low and the
electric mobility is high. The characteristic field for this intrinsic
magnetoresistivity is one magnetic quantum flux $\phi_{0}=h/2e$ per
Fermi wave length area $\pi\lambda_{f}^{2}$. For a density $\varrho=\varrho_{0}\times10^{16}/cm^{3}$,
the field is about $2B_{F}\approx2.92\varrho_{0}^{2/3}$T. It increases
two orders if the density changes three orders. For $\varrho_{0}=10^{3}$,
$2B_{F}\approx292$T. To have an observable magnetoresistivity, the
carrier density should be lower than $\varrho=10^{19}/cm^{3}$. In
fact it has been observed that the magnetoresistance in n-doped germanium
is enhanced with lowing the concentration of impurity \cite{Glicksman1957}
, which is possibly related to the present intrinsic mechanism of
magnetoresistance. Recent discovered Weyl and Dirac semimetals \cite{Armitage2018}
may provides samples with low carrier density and high mobility as
the Fermi level is expected to cross near the Weyl nodes, which are
good candidates for measuring the intrinsic effect.

Finally it is worthy of pointing out that the origin of the intrinsic
negative magnetoresistivity is apparently different from that from
chiral anomaly of Weyl fermions. The chiral anomaly occurs for massless
Weyl fermions in the presence of both electric and magnetic field,
and should be absent for massive fermions. The intrinsic negative
magnetoresistivity persists for either massless or massive Dirac fermions.
Meanwhile the large positive transverse magnetoresistivity is also
irrelevant with the mechanism of chiral anomaly as the electric field
is perpendicular to the magnetic field.

This work was supported by the Research Grants Council, University
Grants Committee, Hong Kong under Grant No. 17301116 and C6026-16W. 

H.W.W. and B.F. contributed equally to this work.

\bibliographystyle{apsrev}

\begin{thebibliography}{localization}
\bibitem{Abrikosov-book} A. A. Abrikosov, Fundamentals of the Theory of Metals (NorthHolland, Amsterdam, 1988).

\bibitem{Pippard1989a} A. B. Pippard, Magnetoresistance in metals, vol. 2 (Cambridge
University Press, Cambridge, 1989).

\bibitem{Armitage2018}N. Armitage, E. Mele, and A. Vishwanath, Rev. Mod. Phys. {\bf90}, 015001 (2018).

\bibitem{Lu2017} H. Z. Lu and S. Q. Shen, Front. Phys. {\bf12}, 127201 (2017).

\bibitem{Kim2013} H. J. Kim, K. S. Kim, J. F. Wang, M. Sasaki,N. Satoh, A. Ohnishi, M. Kitaura, M. Yang, and L. Li, Phys. Rev. Lett. {\bf111}, 246603 (2013).

\bibitem{Xiong2015} J. Xiong, S. K. Kushwaha, T. Liang, J. W. Krizan, M. Hirschberger, W. Wang, R. Cava, and N. Ong, Science {\bf350}, 413 (2015).

\bibitem{Li2016} Q. Li, D. E. Kharzeev, C. Zhang, Y. Huang, I. Pletikosic, A. V. Fedorov, R. D. Zhong, J. A. Schneeloch, G. D. Gu, and T. Valla, Nat. Phys. {\bf12}, 550 (2016).

\bibitem{Li2015a} C. Z. Li, L. X. Wang, H. Liu, J. Wang, Z. M. Liao, and D. P. Yu, Nat. Commun. {\bf6}, 10137 (2015).

\bibitem{Li2015} H. Li, H. He, H. Z. Lu, H. Zhang, H. Liu, R. Ma, Z. Fan, S. Q. Shen, and J. Wang, Nat. Commun. {\bf7}, 10301 (2016).

\bibitem{Huang2015} X. Huang, L. Zhao, Y. Long, P. Wang, D. Chen, Z. Yang, H. Liang, M. Xue, H. Weng, Z. Fang, et al., Phys. Rev. X {\bf5}, 031023 (2015).

\bibitem{Zhang2016a} C. L. Zhang, S. Y. Xu, I. Belopolski, Z. Yuan, Z. Lin, B. Tong, G. Bian, N. Alidoust, C. C. Lee, S. M. Huang, et al., Nat. Commun. {\bf7}, 10735 (2016).

\bibitem{Wang2016a}Z. Wang, Y. Zheng, Z. Shen, Y. Lu, H. Fang, F. Sheng, Y. Zhou, X. Yang, Y. Li, C. Feng, et al., Phys. Rev. B {\bf93}, 121112 (2016).

\bibitem{Zhang2017}C. Zhang, E. Zhang, W. Wang, Y. Liu, Z. G. Chen, S. Lu, S. Liang, J. Cao, X. Yuan, L. Tang, et al., Nat. Commun. {\bf8}, 13741 (2017).

\bibitem{Wang2012} J. Wang, H. Li, C. Chang, K. He, J. S. Lee, H. Lu, Y. Sun, X. Ma, N. Samarth, S. Shen, et al., Nano Res. 5, 739 (2012).

\bibitem{He2013} H. He, H. Liu, B. Li, X. Guo, Z. Xu, M. Xie, and J. Wang, Appl. Phys. Lett. {\bf103}, 031606 (2013).

\bibitem{Wiedmann2016} S. Wiedmann, A. Jost, B. Fauque, J. van Dijk, M. Mei-jer, T. Khouri, S. Pezzini, S. Grauer, S. Schreyeck, C. Brune, et al., Phys. Rev. B {\bf94}, 081302 (2016).

\bibitem{Assaf2017} B. Assaf, T. Phuphachong, E. Kampert, V. Volobuev, P. Mandal, J. Sanchez-Barriga, O. Rader, G. Bauer, G. Springholz, L. de Vaulchier, et al., Phys. Rev. Lett. {\bf119}, 106602 (2017).

\bibitem{Adler1969} S. L. Adler, Phys. Rev. {\bf177}, 2426 (1969).

\bibitem{Bell1969} J. S. Bell and R. Jackiw, Il Nuovo Cimento A  {\bf60}, 47 (1969).

\bibitem{Nielsen1983} H. B. Nielsen and M. Ninomiya, Phys. Lett. B {\bf130}, 389 (1983).

\bibitem{Son2013} D. T. Son and B. Z. Spivak, Phys. Rev. B {\bf88}, 104412 (2013).

\bibitem{Burkov2014a} A. A. Burkov, Phys. Rev. Lett. {\bf113}, 247203 (2014).

\bibitem{Burkov2015} A. A. Burkov, J. Phys. Condens. Matter {\bf27}, 113201 (2015).

\bibitem{Goswami2015} P. Goswami, J. H. Pixley, and S. Das Sarma, Phys. Rev. B {\bf92}, 075205 (2015).

\bibitem{Gorbar2014} E. V. Gorbar, V. A. Miransky, and I. A. Shovkovy, Phys. Rev. B {\bf89}, 085126 (2014).

\bibitem{Gao2017} Y. Gao, S. A. Yang, and Q. Niu, Phys. Rev. B {\bf95}, 165135 (2017).

\bibitem{Andreev2018} A. V. Andreev and B. Z. Spivak, Phys. Rev. Lett. {\bf120}, 026601 (2018).

\bibitem{Murakami2007} S. Murakami, New J. Phys. {\bf9}, 356 (2007).

\bibitem{Burkov2016} A. A. Burkov, Nat. Mater. {\bf15}, 1145 (2016).

\bibitem{Yan2017} B. Yan and C. Felser, Annu. Rev. Condens. Matter Phys. {\bf8}, 337 (2017).

\bibitem{Kane1957} E. O. Kane, J. Phys. Chem. Solids {\bf1}, 249 (1957).

\bibitem{Shen2012} S. Q. Shen, Topological insulators (Springer Nature, Singapore, 2017), 2nd ed.

\bibitem{Shen-11spin} S. Q. Shen, W. Y. Shan, and H. Z. Lu, SPIN {\bf01}, 33 (2011).

\bibitem{Zawadzki-17jpcm} W. Zawadzki, J. Phys. Condens. Matter {\bf29}, 373004 (2017).

\bibitem{Chang2008} M. C. Chang and Q. Niu, J. Phys. Condens. Matter {\bf20}, 193202 (2008).

\bibitem{Xiao2010} D. Xiao, M. C. Chang, and Q. Niu, Rev. Mod. Phys. {\bf82}, 1959 (2010).

\bibitem{Streda1982}  P. Streda, J. Phys. C: Solid State Phys. {\bf15}, L1299 (1982).

\bibitem{Note-on-SM} See Supplemental Material for the calculation details, which
includes Refs. \cite{Gorbar2014,Streda1982,Shoenberg1984,Klier2017,Shen2005,Magnus1943,Fu2017,Elizalde1994,Bastin1971,Wang2016}

\bibitem{Onsager1931} L. Onsager, Phys. Rev. {\bf37}, 405 (1931).

\bibitem{Shoenberg1984} D. Shoenberg, Magnetic Oscillations in Metals (Cambridge University Press, Cambridge, 1984).

\bibitem{Abrikosov1998} A. A. Abrikosov, Phys. Rev. B {\bf58}, 2788 (1998).

\bibitem{Lu2015} H. Z. Lu, S. B. Zhang, and S. Q. Shen, Phys. Rev. B {\bf92}, 045203 (2015).

\bibitem{Zhang2016} S. B. Zhang, H. Z. Lu, and S. Q. Shen, New J. Phys. {\bf18}, 053039 (2016).

\bibitem{Klier2017} J. Klier, I. V. Gornyi, and A. D. Mirlin, Phys. Rev. B {\bf96}, 214209 (2017).

\bibitem{Burkov2017} A. A. Burkov, Phys. Rev. B {\bf96}, 041110 (2017).

\bibitem{Nandy2017} S. Nandy, G. Sharma, A. Taraphder, and S. Tewari, Phys. Rev. Lett. {\bf119}, 176804 (2017).

\bibitem{Li2017} H. Li, H.W. Wang, H. He, J. Wang, and S. Q. Shen, arXiv:1711.03671 (2017).

\bibitem{Wu2017} M. Wu, G. Zheng, W. Chu, W. Gao, H. Zhang, J. Lu, Y. Han, J. Yang, H. Du, W. Ning, et al., arXiv:1710.01855 (2017).

\bibitem{Kumar2017} N. Kumar, C. Felser, and C. Shekhar, arXiv:1711.04133 (2017).

\bibitem{Glicksman1957} M. Glicksman, Phys. Rev. {\bf108}, 264 (1957).

\bibitem{Shen2005} S. Q. Shen, Y. J. Bao, M. Ma, X. C. Xie, and F. C. Zhang, Phys. Rev. B {\bf71}, 155316 (2005).

\bibitem{Magnus1943} W. Magnus and F. Oberhettinger, Formeln und Satze fur die Speziellen Funktionen der Mathematischen Physik (Springer, Berlin, 1943).

\bibitem{Fu2017} B. Fu, W. Zhu, Q. Shi, Q. Li, J. Yang, and Z. Zhang, Phys. Rev. Lett. {\bf118}, 146401 (2017).

\bibitem{Elizalde1994} E. Elizalde, S D Odintsov, A Romeo, A A Bytsenko, and S Zerbini, Zeta regularization techniques with applications (World Scientific, Singapore, 1994).

\bibitem{Bastin1971} A. Bastin, C. Lewiner, O. Betbeder-Matibet, and P. Nozieres, J. Phys. Chem. Solids {\bf32}, 1811 (1971).

\bibitem{Wang2016} C. M. Wang, H. Z. Lu, and S. Q. Shen, Phys. Rev. Lett. {\bf117}. 077201 (2016).

\end{thebibliography}

\end{document}